\magnification=\magstep1
\vsize 23 true cm
\hsize 16 true cm
\parindent 0pt
\parskip 8pt
\def\ot{\varpi}
\centerline{\bf On Stationary, Self-Similar Distributions
}
\centerline{\bf of a Collisionless, Self-Gravitating, Gas}
\vskip  3 true cm
\centerline{\bf R.N. Henriksen and Lawrence M. Widrow}
\medskip
\centerline{\bf Department of Physics}
\medskip
\centerline{\bf Queen's University, Kingston, Canada}
\bigskip
\line{December 14, 1994\hfill}
\vskip .75in
\centerline{\bf Abstract}
\bigskip
We study systematically stationary solutions to the coupled Vlasov and Poisson
equations which have `self-similar' or scaling symmetry in phase space.
In particular, we find analytically {\it all} spherically symmetric distribution
functions where the mass density and gravitational potential are strict
power laws in $r$, the distance from the symmetry point.  We treat as special
cases, systems built from purely radial orbits and systems that are
isotropic in velocity space.  We then discuss systems with 
arbitrary velocity space anisotropy finding a new and very
general class of distribution functions.  These
distributions may prove useful in modelling galaxies.  
Distribution functions in cylindrical and
planar geometries are also discussed.  Finally, we study spatially spheroidal
systems that again exhibit strict power-law behaviour for the density and
potential and find results in agreement with results published recently.

\vfill\eject
\vskip 2 true cm
\line{\bf 1~~ Introduction \hfill}
\bigskip

Star clusters, dark matter galactic halos, and clusters of galaxies are essentially 
collisionless self-gravitating systems obeying the coupled Poisson-Vlasov
equations and therefore equilibrium solutions to these equations are of
great importance.  While substantial progress has been made through numerical 
simulation, there has also been a persistent school of 
analysis that seeks to obtain exact, analytic results, particularly in asymptotic or 
`stationary' limits. 

One branch of this latter school has studied the evolution of collisionless
matter in an expanding universe (Fillmore \& Goldreich 1984; Bertschinger 1985;
Gurevich \& Zybin 1988, 1990; Ryden 1993).  The solutions are time dependent of necessity but 
show steady or at least adiabatic behaviour at late times.
The treatments make use of an intuitive self-similar symmetry
which can seem rather ad hoc (even if exceedingly 
clever) and so, difficult to access and generalize.  Moreover, there are some 
remaining inconsistencies in the various published results. Our intention 
here is to study such collisionless self-similarity in a simple and systematic 
way, beginning with strictly stationary examples.

Self-similar symmetry has also been found in the study of collisional systems such as 
the cores of globular clusters (e.g. Lynden-Bell 1967; Lynden-Bell and 
Eggleton 1980; Inagaki and Lynden-Bell 1983, 1990). These systems  allow the 
study of the evolution towards core-halo configurations but never yield true 
thermodynamic equilibria in the form say of stationary power law behaviour. 
Nevertheless in various intermediate (spatially and temporally) 
stages simple power law do appear and may in fact correspond to the 
solutions we find for stationary collisionless systems. 
It is well known in the hydrodynamic literature (e.g. Barenblatt and 
Zel'dovitch 1972, hereafter BZ) that self-similar solutions arise as intermediate asymptotics 
between boundaries, and it is probably for this reason that they are found  in 
the pre and post collapse phases of the collisional systems. One might speculate that 
they are as close to `equilibrium' as such systems get.  In addition to globular clusters,
these systems may arise in the intermediate stages of collapse of a 
star cluster to a black hole and in the ultimate state of collisionless dark 
matter halos.      

In this paper we study systematically the family of distributions that are 
exactly stationary and that possess both precise geometric (usually 
spherical) and scaling symmetries.  These distributions are solutions
to the coupled Vlasov and Poisson equations:
$$\sum_{i=1}^3\left (
{\partial H\over\partial v_i}{\partial f\over\partial x_i}~-~
{\partial H\over\partial x_i}{\partial f\over\partial v_i}
\right )~=~0\eqno(1.1)$$
$$\nabla^2\Phi~=~4\pi G\int f d^3 v~.\eqno(1.2)$$
Here $f$ is the usual mass distribution function, $\Phi$ is the gravitational
potential, $H$ is the Hamiltonian and $v_i,\, x_i$ are canonical pairs.
The solutions we find include the intermediate 
asymptotic limits 
of the relevant preceding studies. They overlap directly with the recent 
work of Evans (1994, hereafter E94) on ``power law galaxies" and to a lesser extent 
with the `$\eta$' models of Tremaine et al. (1994). These latter models are 
similar in spirit to our own but are not strictly scale free and so 
prove to be rather more complicated to express and use. Nevertheless they 
share the observationally important property of a cusped central density
and possess the substantial advantage of having finite mass.  

The power-law galaxies of E94 (see also Evans and de Zeeuw 
1994, hereafter EdZ) are scale invariant and therefore simple to use.
They were constructed to provide a family of
versatile models of galaxies that could be used to analyse observable
properties of galaxies such as line profiles from stellar
absorption-line spectra.  Our approach is more systematic and in fact,
for the case of spherical symmetry with velocity space anisotropy,
yields a more general class of solutions.
We mostly leave applications such as those found in van der Marel \& Franx (1993),
E94 and EdZ for future
publications.

We also use this paper to introduce to the astronomical community
a systematic method of classifying scaling or self-similar  
symmetry first given by Carter and Henriksen (1991, hereafter CH) in a 
rather formal style. The technique is easy to use and yields the
reduced equations (with manifest scaling symmetry) in a standard form.  In addition,
the method includes the various scaling symmetries that produce irrational
power laws formerly referred to as scaling symmetries of the `second' kind (e.g. BZ).

\bigskip\bigskip

\line{\bf 2~~ Radial Orbits in Spherical Symmetry;\hfill}
\line{\bf An Introduction to the 
CH Analysis of the Vlasov-Poisson Equations\hfill}

\bigskip

We introduce the canonical distribution function $F$ where 

$$f({\bf r},{\bf v})\equiv {F(r,v_r)\over 4\pi r^2}
\delta\left (v_\theta\right )\delta\left( v_\phi\right )~.\eqno(2.1)$$

The Vlasov and Poisson equations become 

$$v_r\partial_r F - \partial_r\Phi\partial_{v_r}F=0~,\eqno(2.2)$$

$$\partial_r(r^2\partial_r\Phi)=G\int F\,dv_r~.\eqno(2.3)$$

Following CH we seek a self-similar or scaling 
symmetry in phase space by requiring that the distribution function satisfy
the equation

$${\cal L}_{\bf k} F=0\eqno(2.4)$$

where

$${\cal L}_{\bf k}~\equiv~
k^j\partial_j\equiv \delta r\partial_r + \nu v_r \partial_{v_r}\eqno(2.5)$$

is the Lie derivative with respect to the (phase space) vector operator ${\bf k}$.
In other words, ${\bf k}$ is the scaling direction in phase space.
It is convenient in these formulae to imagine that $r$ and $v_r$ are scaled 
interms of fiducial values (not to be confused with real constant lengths).

Equation (2.4) holds 
so long as $F=F(\zeta)$ where $\zeta\equiv r^{\left(\nu/\delta\right )}/v_r$.  The 
dimensionless real number $\delta/\nu$ gives the similarity `class' of the symmetry 
in the sense of CH and it is generally fixed only by boundary conditions or 
by the dimensions of conserved quantities.
 
Our first step is to choose new phase space coordinates to replace $r$ and $v_r$.  
Following CH we define a new `radial' coordinate $R(r)$ 
such that 

$$ {\cal L}_{\bf k}~=~k^j\partial_j~\equiv~ \partial_R~.\eqno(2.6)$$

$k^j\partial_j R=1$ and we obtain the transformation laws 

$$r\mid \delta\mid~=~e^{\delta R}~,\eqno(2.7)$$

and 
 
$${dR\over dr}~=~e^{-\delta R} sgn(\delta)~.\eqno(2.8)$$

The orthogonal invariant or self-similar variable $X$ (in the notation of CH)
satisfies $k^j\partial_jX=0$.  In this simple example 
$X$ is an arbitrary function of $\zeta$.  Without loss of generality, we take it to
be

$$X~\equiv~\mid\delta\mid^{-(\nu/\delta )}/\zeta~=~
v_r e^{-\nu R}.\eqno(2.9)$$

As of yet, there are no restrictions on the solutions;
($X$, $R$) is merely a new coordinate system.
The restrictions appear only subsequently when
we impose the invariance of various quantities under the action of ${\bf k}$.

While the scaling of dimensional quantities under the action ${\bf k}$ is uniquely 
determined, the form is hardly 
standard and varies from author to author.
CH present an alternative
analysis introducing a `dimensional' algebra in an appropriate `dimension 
space'. In the current example it is sufficient to choose a 
dimension space consisting of length, velocity and mass. There is then a 
three parameter multiplicative rescaling group with elements ${\bf A}\equiv 
(e^\delta, e^\nu, e^\mu)$ that describes the scalings of dimensional
quantities.  Alternatively, if we consider changes in the logarithms of 
these quantities, we have an additive rescaling group with elements
${\bf a}=(\delta,\nu, \mu)$. The vector components of ${\bf A}$ or ${\bf a}$
correspond to the scaling in length, velocity and mass respectively. 

Each dimensional quantity $\Psi$ in the problem has its dimensions represented 
by a dimensionality 
(co)vector ${\bf d_\Psi}$ in the dimension space, and the change in the logarithm 
of the quantity is given by (CH) 

$${\cal L}_{\bf k}\Psi~=~
\partial_R\Psi~=~
({\bf d}_\Psi \cdot {\bf a})\Psi~.\eqno(2.10)$$

Strictly speaking, the vector ${\bf k}$ should correspond to the rescaling 
vector ${\bf a}$.  In other words, we should replace equation (2.5) with 

$$k^j\partial_j~=~\delta r\partial_r~+~\nu v_r\partial_{v_r}~+~\mu m\partial_m~.\eqno(2.11)$$

However as we will soon see, the invariance of $G$ under rescaling implies that 
mass rescaling is not independent of length and velocity rescaling.  This allows us to 
reduce our scaling algebra element to ${\bf a}=\left (\delta,~\nu\right )$.

The dimensional quantities in 
the current problem $F$, $\Phi$ and $G$ have the following 
dimensionality 
covectors in the chosen dimension space {\it (length, velocity, mass)};

$$\eqalignno{{\bf d}_F&=(-1,-1,1)~,\cr    
             {\bf d}_\Phi&=(0,2,0)~,& (2.12)\cr
             {\bf d}_G &=(1,2,-1)~.\cr}$$
             
The requirement that $G$ be invariant 
under the rescaling group action implies
${\bf a}\cdot{\bf d_G}=0$, or on performing the scalar product (direct multiplication
   since ${\bf d}$ is a covector)

$$\mu=2\nu+\delta~.\eqno(2.13)$$

Consequently the dimension space may be reduced to the sub-space of 
$(length,velocity)$ wherein the 
rescaling group element becomes ${\bf a}=(\delta,\nu)$ and 

$$\eqalignno{{\bf d}_F&=(0,1)~,\cr    
             {\bf d}_\Phi&=(0,2)~.& (2.14)\cr}$$                  
  
The sub-space algebra element ${\bf a}$ now corresponds to our choice of the scaling 
vector ${\bf k}$.
 
The scaling symmetry can be imposed on $F$ and $\Phi$ by equation (2.10) which,  
with equation (2.14) requires 

$$\eqalignno{F(X,R)&~=~\overline{F}(X) e^{\nu R}~,\cr
              \Phi(X,R)&~=~\overline{\Phi}(X) e^{2\nu R}~.&(2.15)\cr}$$ 

These equations can be simplified further by noting that
at fixed $r$ the potential is independent of $v_r$ or equivalently

$$\overline{\Phi}~=~constant~.\eqno(2.16)$$

Substituting into the Vlasov and Poisson equations 
yields

$$ {d\ln\overline{F}\over d\ln X}={X^2\over X^2+2\overline{\Phi}}~,\eqno(2.17)$$
  
and 

$$2\left ({\delta\over\nu}+2\right )\overline{\Phi}~=~G\left ({\delta\over\nu}\right )^2
\int\, \overline{F}\, dX~.\eqno(2.18)$$
   
It is now plain that the class of solutions depends only on the ratio $\delta
/\nu$ and so without loss of generality, we set $\nu=1$.
  
Equation (2.17) now integrates to yield $\overline{F}=C \mid 
X^2+2\overline{\Phi}\mid^{(1/2)}$ and therefore by equations (2.9) and (2.15)

$$F~=~C \mid v_r^2 +2\Phi\mid^{(1/2)}~\equiv ~C
\mid 2E\mid^{(1/2)}.\eqno(2.19)$$

Here $C$ is a normalization constant given below.
The potential is given by $\Phi=\overline{\Phi}e^{2R}=\overline{\Phi}
\left (\mid\delta\mid r\right )^{2/\delta}$.  It is convenient to introduce
an explicit fiducial length $a$.  The potential can then be written as

$$\Phi = \Phi_a\left ({r\over a}\right )^{2/\delta}\eqno(2.20)$$

where $\Phi_a$ has units of $velocity^2$ and replaces $\overline\Phi$.
We also find from (2.18) that

$$C~=~{ 2\mid\delta+2\mid\over\pi\delta^2 G}~. \eqno(2.21)$$ 

For bound solutions, $\overline\Phi$ (or equivalently $\Phi_a$) must be negative
and therefore $\delta<-2$.  Formally $\delta>0$ and 
$\overline{\Phi}>0$ also give an inwardly directed gravitational force, 
but now there is no natural cut-off in equation (2.18) and the whole integral does not
converge.

The mass density obeys a simple power law:

$$\rho\propto  r^{(2/\delta-2)},\eqno(2.22)$$

and $\delta<-2$  {\it implies that the density law lies in the range} 
$-2>{2\over\delta}-2>-3$.  In addition, the velocity moments of the
distribution function also follow simple power laws:
   
$$\overline{v_r^{2n}} \propto r^{2n/\delta}\eqno(2.23) $$

indicating that velocities increase without limit as one moves towards the center
of the distribution.

It is interesting that this distribution function has the same functional dependence
on energy as a Plummer model 
with $n=5/2$.  However, the density does not satisfy the Lane-Emden equation 
since it is `cusped' at the centre and so does not have the right boundary 
conditions (e.g. Binney \& Tremaine 1987; hereafter BT).  It appears to be the distribution
function corresponding to the asymptotic
solutions found by Fillmore and Goldreich (1984).  Indeed,
although their work was done in the context of an expanding universe,
their solutions in the case of spherical symmetry turn out to
be time-independent (see their equation (39)).  Figure 1 shows a contour map of the 
distribution function for $\delta=-17$.  This value of $\delta$ is chosen
so that our Figure 1 corresponds to their Figure 10.  Contours give equal probability
density in the correct phase space and so the integral of the
contour value times the area element for any region of the plot gives
the total number of particles in that region of phase space.  In addition particle orbits
are coincident with these contours and in fact Fillmore and Goldreich (1984)
construct their plots by calculating particle orbits.

We are left with the free parameter $\delta$ in the range $(-2,-\infty)$. 
How might this be determined? It is clear from the discussions of earlier 
papers that in general it appears as a sort of eigenvalue when the solution 
is seen to arise from more general initial conditions. However another manner 
in which it may be fixed is to require additional global invariants.  For 
example, if we introduce a characteristic mass, which may be either 
the mass of each particle or that of a central massive object (or both in 
the sense that once there is one fixed mass we can measure all others in terms 
of it). Then there will be no mass scaling in the scaling algebra which 
means $\mu=0$. From this and equation (2.13) we see that $\delta=-2$ ! This 
corresponds to a central point mass surrounded by massless particles that 
nevertheless are distributed in energy space according to equation (2.19).  
   
Having illustrated our method in detail in this simple example with radial 
orbits, we proceed to give our results briefly for other cases of interest.
    
\bigskip\bigskip

\line{\bf 3~~Isotropic Orbits in Spherical Symmetry \hfill }
      
\bigskip

Here, the full 3D distribution function has
 the functional form $f=f(r,v)$, 
where $v^2=v_r^2+v_\theta^2+v_\phi^2$. The Vlasov and Poisson equations 
are now respectively ($v_r\ne 0$)

  $$\eqalignno{v\partial_r f-(\partial_r\Phi)\partial_{v} f&~=~0,& (3.1)\cr
               {1\over r^2}\partial_r(r^2\partial_r\Phi)&~=~4\pi G\int\,4\pi v^2 
               f\, dv~.&(3.2)\cr}$$

The analysis proceeds as above and in fact equations (2.6), (2.7) and (2.8) apply here as well 
while equations (2.5) and (2.9) hold with $v_r$ replaced by $v$. 
The dimension space is again {\it (length, velocity, mass)}.  The dimensionality
covector of $f$
becomes ${\bf d}_f=(-3,-3,1)$ while those of $G$ and $\Phi$ remain as in 
equation (2.12). The invariance of $G$ requires equation (2.13) as before so 
that the dimensionality covectors in the subspace of $(length,velocity)$ are 

  $$\eqalignno{{\bf d}_f&~=~(-2,-1)~,\cr
               {\bf d}_\Phi&~=~(0,2)~,&(3.3)\cr}$$

whence following equation (2.10) 

  $$\eqalignno{f&~=~\overline{f}(X)e^{-(2\delta +\nu)R}~,\cr
               \Phi&~=~\overline{\Phi}(X)e^{2\nu R}~.&(3.4)\cr}$$

We will set $\nu=1$ in the sequel to be consistent with the notation in Section 2.
Again, things simplify as $\overline\Phi = constant$.  
Substituting into equations (3.1) and (3.2) yields

  $${ d\ln\overline f \over dX}~=~-{(2\delta +1)X\over X^2+2\overline\Phi}\eqno(3.5)$$
 
and  

  $$2(\delta+2)\overline\Phi~=~(4\pi)^2G\int_0^{\sqrt{-2\overline\Phi}}\,X^2\,
  \overline f(X)\, dX~.\eqno(3.6)$$

Equation (3.5) easily integrates and we find

  $$\overline f~=~C\mid X^2+2\overline\Phi\mid^{-(\delta+1/2)}~.
  \eqno(3.7)$$

It becomes clear from this last equation that for a bound system ($\overline
\Phi <0$) once again $\delta<-2$.  In this case however, unbound solutions do
exist since the integral in (3.6) converges for $\delta>1$. 
  
We may use equations (3.4) and (2.9) (with $v$ in place of $v_r$) to write the 
solution for the distribution function as 

  $$f~=~C\mid v^2+2\Phi\mid^{-(\delta+1/2)}\equiv C\mid 2E\mid^{-(\delta+
  1/2)},\eqno(3.8)$$

and that for the potential as 

  $$\Phi=\overline\Phi\left ( \mid \delta \mid r \right )^{2/\delta}\equiv
\Phi_a\left ({r\over a}\right )^{2/\delta}.\eqno(3.9)$$

As before, $a$ is a fiducial length and $\Phi_a$ is a constant with units
$velocity^2$ that replaces $\overline\Phi$.  The normalization constant $C$ 
is determined from the Poisson equation:

$$C~=~\left \{ {-\left (1+\delta/2\right )
\left (1-\delta\right )\Gamma\left (1-\delta\right )\over
2\pi^{5/2}\delta^2\Gamma\left (1/2-\delta\right )}\right \}
{\left (-2\Phi_a\right )^\delta\over G a^2}~.\eqno(3.10)
$$

It is interesting to note that
the corresponding density law has
the same functional form as in the pure radial case discussed above.  Moreover,
$\overline{v^{2n}}\propto r^{2n/\delta}$.
  
Once again, at the cost of admitting a singular potential and density we have found 
a large class of isotropic analytic solutions. The distribution function of equations (3.8)
and (3.9) is 
much like the Plummer model (e.g. BT, p223), but the potential member of the 
pair is singular. The stability theorems of Antonov and of Doremus, Feix and 
Baumann (as cited in BT section 5.2) seem to imply moreover that this
distribution is stable.  The velocity dispersion for both radial and
isotropic cases increases towards 
towards the centre of the distribution even though the mass at the centre is zero.
(The mass inside radius $r$ diverges $r\to \infty$ in common with the standard `isothermal' 
model.)  These solutions therefore might be
useful models for star clusters at the centers
of active galaxies. For example, if a detailed mapping of the velocity dispersion 
in M87 were to show a slower than Keplerian increase with decreasing $r$, then 
star cluster models of this type might provide an alternative to 
the massive black hole theory. 
  
We note that once again the limit $\delta=-2$ corresponds to an invariant
central mass surrounded by massless particles. This is the 
Keplerian, and possibly the central black hole, limit.    

\bigskip\bigskip

\line{\bf  4~~Systems with Cylindrical and Planar Symmetry\hfill}

\bigskip

It is straightfoward to extend the previous analysis to systems with
cylindrical and planar symmetry.  Such solutions might provide insight
into the filaments and sheets seen both in N-body simulations and in
large-scale galaxy surveys.  Our solutions should correspond to those
found in Fillmore and Goldreich (1984) though our solutions are
stationary and do not include cosmological expansion.

For systems with cylindrical symmetry, we choose the canonical distribution
function

$$f({\bf r},{\bf v})~\equiv~ { F (\ot,v_\ot )\over 2\pi\ot}
\delta\left (v_\theta\right )\delta\left( v_z\right )\eqno(4.1)$$

where $\ot$ is the distance from the axis of symmetry.
The Vlasov and Poisson equations are then

$$v_\ot\partial_\ot F - \partial_\ot\Phi\partial_{v_\ot}F=0~,\eqno(4.2)$$

$$\partial_\ot(\ot\partial_\ot\Phi)=G\int F\,dv_\ot~.\eqno(4.3)$$

We find the family of solutions

$$f~\propto~|E|^{(1-\delta)/2}~~~~~~~~~~~~\Phi~\propto~\ot^{2/\delta}~.\eqno(4.4)$$

The $mass/length$ inside a radius $\ot$ is $\mu(\ot)\propto\ot^{2/\delta}$
while the density is $\propto\ot^{2/\delta-2}$.  If we require that
the density decrease with increasing $\ot$ (but allow for a cusped
density law at the origin) and also require that there be no central
mass concentration ($\mu(\ot)\to 0$ for $\ot\to 0$) then $\delta$ is constrained
to be greater than $1$.

For planar symmetry, we choose

$$f({\bf r},{\bf v})\equiv F(z,v_z)
\delta\left (v_x\right )\delta\left( v_y\right )~.\eqno(4.5)$$

The Vlasov and Poisson equations are then

$$v_z\partial_\ot F - \partial_z\Phi\partial_{v_z}F=0~,\eqno(4.6)$$

$$\partial_z^2\Phi=G\int F\,dv_z~.\eqno(4.7)$$

In this case, the family of solutions is

$$f~\propto~|E|^{(1-2\delta)/2}~~~~~~~~~~~~\Phi~\propto~z^{2/\delta}~.\eqno(4.8)$$

$1<\delta <2$ is enough to insure that the density will decrease with increasing 
$z$ and the mass per unity area will vanish for $z\to 0$.

\bigskip\bigskip

\line{\bf 5~~ Anisotropic Orbits in Spherical Symmetry\hfill}

\bigskip

This example is more challenging then the others treated in this paper
and the results are somewhat surprising.  The distribution function depends on 
three phase space coordinates
$(r,v_r,j^2)$ where $j^2\equiv r^2(v_\theta^2+v_\phi^2)$
is the square of the transverse angular momentum (Fujiwara 1983). The Vlasov and 
Poisson equations become respectively

$$\eqalignno{v_r\partial_r f +\left({j^2\over r^3}-\partial_r\Phi\right)
\,(\partial_{v_r}f)&=0,&(5.1)\cr
\partial_r(r^2\partial_r\Phi)-4\pi^2 G\,\int\, dj^2\int\,f(r,v_r,j^2)\,dv_r&=0.
&(5.2)\cr}$$

The dimension space is taken to be 
{\it (length, velocity,(angular momentum)$^2$, mass)}, with the 
scaling algebra vector ${\bf a}=(\delta,\nu,\lambda,\mu)$. 
However the invariance of $G$ allows us to work in the reduced scaling space
{\it (length, velocity, (angular momentum)$^2$)}.  Moreover we know that 
we can set one of the scale factors equal to unity since only the ratios have 
physical significance, and we choose this to be $\nu=1$ as above. 
The symmetry we seek can therefore be written in the simplified form 

$$k^j\partial_j~\equiv~ r\delta\partial_r +v_r\partial_{v_r} +\lambda j^2\partial_
{j^2}~.\eqno(5.3)$$

As before we choose $R$ to lie along this direction so that equations (2.6), (2.7) 
and (2.8) continue to apply. 

There must now be {\it two} invariant coordinates ${X^i}$ orthogonal to $R$ 
which satisfy $k^j\partial_jX^i=0$. These are linear, first-order partial 
differential equations that integrate easily to give 
$$X^i=\overline{X}^i(\zeta)\times v_r\,e^{-R},\eqno(5.4)$$
where $\overline{X}^i$ are arbitrary functions of $\zeta\equiv v_r\,
j^{-2/\lambda}$. It is convenient to choose $\overline{X}^{(1)}=1$ and 
$\overline{X}^{(2)}=1/\zeta$ so that 
$$X^{(1)}\equiv X= v_r\,e^{-R}\eqno(5.5)$$
and 
$$X^{(2)}\equiv Y^{(1/\lambda)}=j^{(2/\lambda)}\,e^{-R}~.\eqno(5.6)$$

We shall write our equations in terms of $R,~X$, and $Y$.

In the reduced dimension space the quantities have the dimensionality covectors 
$$\eqalignno{{\bf d}_f&=(0,1,-1)~,\cr
{\bf d}_\Phi&=(0,2,0)~,&(5.7)\cr}$$
whence as usual 
$$\eqalignno{f&=\overline{f}(X,Y) e^{(1-\lambda)R}~,\cr
\Phi&=\overline{\Phi}(X,Y) e^{(2R)}~.&(5.8)\cr}$$
The potential at fixed $r$ should be independent of $v_r$ 
and $j^2$ and we impose $\overline\Phi=constant$.

The Vlasov and Poisson equations now become 
$$\eqalignno{-(2\delta+1)\overline f-X\partial_X\overline f-2(1+\delta)Y\partial
_Y\overline f +{1\over X^2}(\delta^3 Y-2\overline\Phi)X\partial_X\overline f
&=0,&(5.9)\cr
2\overline\Phi\left({1\over\delta}+{2\over\delta^2}\right)
-4\pi^2G\,\int\,dY\,\int\,dX\,\overline{f}(X,Y)&=0.&(5.10)\cr}$$

where

$$\lambda~=~2(1+\delta)\eqno(5.11)$$

follows from the requirement that these equations be independent of $R$.
We see from equation (5.10) that $\delta<-2$ for the system to be 
bound. 

Equation (5.9) is a quasi-linear, first-order partial differential equation that 
can be solved exactly. The characteristic equations are 
$${d\overline f\over (2\delta+1)\overline f}~=~{dY\over -2(1+\delta)Y}~=~
{X\,dX\over (\delta^3 Y-X^2-2\overline\Phi)}~,\eqno(5.12)$$
the first of which is clearly integrable and gives 
$$\overline f~=~\overline {F}(\xi)\, Y^{-\left(2\delta+1\over 2(1+\delta)\right)}
,\eqno(5.13)$$
where $\xi$ is a function constant on the integral curves of the second equation 
of (5.12). 
This second equation may be integrated by introducing the variables $u$, $s$ 
 such that 
$$s~\equiv~ X^2+2\overline\Phi~,\eqno(5.14)$$
and 
$$\delta^3Y~\equiv~ u+s~.\eqno(5.15)$$

The equation to be integrated is now reduced to 
$$u\,{du\over ds}+(2+\delta)u + (1+\delta)s~=~0~,\eqno(5.16)$$
which is of a type already known to Leibnitz in 1691. The solution is immediate
for $s\ne 0$ by changing the dependent variable to say $y\equiv u/s$. We find 
thereby that the quantity $\xi(X,Y)$ may be taken to be 

$$\xi=Y^{-1}\mid \delta^2Y+X^2+2\overline\Phi\mid^{(\delta + 1)}.\eqno(5.17)$$

The solution is now determined {\it but for the choice of the arbitrary function 
} $\overline{F}(\xi)$.   The constant $\overline\Phi$ is determined 
from equation (5.10) or equivalently 
$$2\overline\Phi\left({1\over\delta}+{2\over\delta^2}\right)
=4\pi^2G\,\int_0^{-2\overline\Phi/\delta^2}\,dY\,\int_{-\sqrt{-2\overline\Phi
-\delta^2Y}}^{\sqrt{-2\overline\Phi-\delta^2 Y}}\,dX\,\overline f.\eqno(5.18)$$
In this formula $\overline{f}$ is used in the form of equation (5.13) and the 
order of integration may be reversed if the integrals exist. 

It is instructive to use equations (5.5), (5.6) and (5.8) to return to physical 
variables whence 

$$\xi=j^{-2}\mid 2\Phi+v^2\mid^{\delta+1}
\equiv j^{-2}\,\,\mid 2E\mid^{\delta+1},\eqno(5.19)$$
and
$$f=j^{-\left(2\delta+1\over\delta+1\right)}\overline{F}(\xi)
\eqno(5.20)$$

where $v^2\equiv v_r^2+{j^2/ r^2}$ is the squared magnitude of the velocity. 
It is remarkable that despite the arbitrary nature of 
$\overline{F}(\xi)$, equations (2.20) and (2.22) continue to give the potential 
and density power laws. The amplitude $\overline{\Phi}$ does depend however 
on $\overline{F}(\xi)$ through equation (5.18) and one must insure that the integrals in
this equation exist.

Equation (5.20) represents a new class of distribution functions
with velocity space anisotropy.
Models with $\overline F(\xi)\propto\xi^\alpha$ correspond to the power-law
spheres discussed in E94.  For these models, the distribution function takes the simple
form

$$f = C j^a\mid E\mid ^b\eqno(5.21)$$

where $a$ and $b$ are related to $\alpha$ and $\delta$ in a simple way.
(Actually, distributions with energy and angular momentum dependence given
by equation (5.21) were first discussed by Camm (1952).  There, the spatial
dependence of the mass density and potential are given by solutions to the
Emden-Fowler equation with appropriate boundary conditions.)
As discussed by E94, it is straightfoward to calculate velocity
moments for these distribution functions.  For example, one can calculate the velocity
space anisotropy
parameter $\beta\equiv 1 - \overline{v_\theta^2}/\overline{v_r^2}$:

$$\beta ~=~ {2\delta + 1\over 2\left (\delta + 1\right )}~+~\alpha\eqno(5.22)$$

Here, $\beta=0$ $\left (\alpha=-\left (2\delta+1\right )/2\left (\delta+1\right )\right )$
corresponds to an isotropic distribution in velocity space,
$\beta=1$ $\left (\alpha=1/2\left (\delta+1\right )\right )$
corresponds to purely radial orbits, and $\beta\to -\infty$ $\left (
\alpha\to -\infty\right )$ corresponds 
to a distribution of purely circular orbits.  Evidently, equation (5.21) describes a 
two-parameter family of distribution functions where one
parameter specifies the density law and the other
parameter specifies the anisotropy in velocity space.
The distribution given in equation (5.20) is more general than this indicating that
entirely different distribution functions can have the same
density law and velocity space anisotropy.  We illustrate this through the 
sequence of figures 2(a)-2(d).  These figures show contour plots of 
distribution functions in velocity space $\left (\left (j/r,~v_r\right )~{\rm space}\right )$
for fixed $r$.  Figs 2(a)-2(c) are power-law spheres described by equation (5.21)
and correspond to an isotropic distribution ($\beta=0$); a distribution constructed
from nearly radial orbits ($\beta=0.9$); and a distribution constructed from
nearly circular orbits ($\beta=-9.0$) respectively.  Fig 2(d) is a composite 
model constructed from the power-law models used in Fig 2(b) and 2(c), i.e., from
nearly radial and nearly circular oribits.  By construction, $\beta=0$ as in the isotropic
case though the actual distribution function is entirely different.

It is interesting to note that
the degeneracy of models discussed here
(density law fixed 
while the functional form (admittedly of a specified argument in $j$ and $E$)
free) is in the opposite sense to the 
degeneracy found in the `Einstein model' where particles move on all possible circular orbits 
in spherical symmetry while the density and potential laws are arbritrary
(Fridman and Polyachenko 1984).

It is plain that a great deal remains to be discovered about these distribution functions,
which have the character of a `post core collapse' (zero central mass) 
stationary state.  For example, for which choices of $\overline F(\xi)$
are they stable?  Fridman and Polyachenko (1984) show that the Camm models are linearly unstable 
for sufficiently large anisotropy, but are stable towards the isotropic limit.
But linear stability is not the same as non-linear stability which has really to
do with the existence or otherwise of asymptotic distributions towards which
$f$ tends.  By studying the stability of our stationary solutions we might
discover a hint as to how to define a useful `entropy' function
that characterizes such asymptotic equilibria.  This latter question has recently been
given new life by the studies of Tremaine et al. (1986), Wiechen et al. (1988)
and Aly (1989, 1993).  This last author has shown for example
that the `softened' Plummer model with $\delta=-4$ in our notation actually attains the minimum
energy subject to a fixed mass and a fixed value of the `entropy'.
If this value is used naively in our models,
one obtains $\rho\propto r^{-5/2}$ and $\Phi\propto r^{-1/2}$.  Finally one might 
ask how closely can a {\it stable} model approach Maxwellian
type distributions.

\bigskip\bigskip

\line{\bf 6~~ Axially Symmetric Solutions with Ellipsoidal\hfill} 
\line{\bf and Hyperboloidal Symmetries\hfill}

\bigskip

In order to make contact with the recent studies of E94 and of EdZ 
we show here how our method yields the strictly scaling 
subset of their axially symmetric solutions in a very direct way. The scaling symmetry in phase 
space for this case is actually simpler than that for the anisotropic 
spherical geometry dealt with above. 
 We use cylindrical coordinates 
${\ot,\phi,z}$ and write the Vlasov equation in the symmetric form 
$$v_\ot\partial_\ot\,f+v_z\partial_z\,f+\left({v_\phi^2\over \ot}-\partial_\ot\,\Phi
\right)\partial_{v_\ot}\,f-{v_\ot v_\phi\over \ot}\partial_{v_\phi}\,f-
\partial_z\Phi\partial_{v_z}\,f=0,\eqno(6.1)$$
while the Poisson equation becomes
$${1\over \ot}\partial_\ot(\ot\partial_\ot\Phi)+(\partial_z)^2\Phi=4\pi G\int\,f
d\,v_\ot\,d\,v_\phi\,d\,v_z\,\, .\eqno(6.2)$$

Following E94 we seek solutions with confocal ellipsoidal symmetry wherein  
$\Phi=\Phi(u)$ where 
$$u^2\equiv \ot^2+{z^2\over q^2}~.\eqno(6.3)$$
We differ from E94 in that we do not introduce a `core' radius into 
the problem, since strictly speaking this would prohibit the existence of 
true geometrically scaling solutions.  It will be clear however from our method that one 
can add a core radius squared to $u^2$ without changing the solution, {\it 
provided 
that it is ignored in the scaling symmetry}. Thus our solutions as well as those 
of E94 do not `really' know of its existence and so we prefer to suppress it.
It is amusing to note also that the substitutions $q^2\rightarrow -q^2$ and 
$u\rightarrow v$ where 
$$v^2\equiv \ot^2-{z^2\over q^2},\eqno(6.4)$$
in our results will convert them to hyperboloidal symmetry in which the equipotential 
surfaces are confocal hyperboloids of one sheet rotated about the $z$ axis. 
Since these extend to infinity it may be that they are of little interest, but 
we note that near $z=0$ they tend to a configuration that might be associated 
with a vertically stratified disc containing a central `hole'. 

We proceed with a direct substitution of $\Phi=\Phi(u)$ into the right hand side
of (6.2) followed by  a collection of terms in equal powers 
of $\ot$ to discover the necessity for a distribution function in the ansatz form 
$$f=f_1(u,v_\ot,v_\phi,v_z)+f_2(u,v_\ot ,v_\phi,v_z)\, \ot^2 .\eqno(6.5)$$
Hence the Poisson equation splits into the two equations 
$$\eqalignno{4\pi G&\int\,f_1\, d\,v_\ot\,d\,v_\phi\,d\,v_z\,={2\Phi'\over u}+
{\Phi''\over q^2},&(6.6a)\cr
4\pi G&\int\,f_2\,d\,v_\ot\,d\,v_\phi\,d\,v_z\,=\left({1\over q^2}-1\right)\left(
{\Phi'\over u^3}-{\Phi''\over u^2}\right),&(6.6b)\cr}$$
where the primes denote total derivatives with respect to $u$. 

The key procedure is the substitution of the ansatz (6.5) into the Vlasov equation 
(6.1) and the subsequent need to satisfy the equation for the coefficients of each 
of the powers of $\ot$ and of $z$. Each such coefficient presents us with simple 
quasi-linear partial differential equations to solve and despite the apparent 
danger of over determining 
the problem, everything in fact works smoothly. Thus from setting the coefficient 
of $z$ equal to zero we learn that 
$$\eqalignno{f_1&~=~f_1\left ({v_z^2\over 2}+\Phi,v_\ot ,v_\phi\right )~,\cr
f_2&~=~f_2\left ({v_z^2\over 2}+\Phi,v_\ot ,v_\phi\right )~.&(6.7)\cr}$$
Then by setting the coefficient of $\ot^{-1}$ equal to zero there follows finally for 
$f_1$
 
$$f_1~=~f_1\left( {v_\ot^2+v_\phi^2+v_z^2\over 2}+\Phi
\right)~\equiv~ f_1(E).\eqno(6.8)$$

The vanishing of the coefficient of $\ot$ yields then directly for $f_2$ 

$$f_2~=~v_\phi^2\,F_2\left(E\right),\eqno(6.9)$$ 

after which the coefficient of the last term in $\ot^3$ vanishes identically.
Consequently we concur with E94 that the necessary ansatz (6.5) is in 
fact of the form 

$$f~=~f_1(E)+\ot^2\,v_\phi^2\,F_2(E)~\equiv~ f_1(E)+j_z^2\,F_2(E)~,\eqno(6.10)$$

although at present $f_1$ and $F_2$ are arbitrary functions of the energy $E$.
The Vlasov equation is now identically satisfied, in accordance with Jeans' theorem 
for stationary solutions.
 
 As in the previous sections we turn now to the imposition of a scaling 
 symmetry in phase space. The Lie derivative will be along the vector 

 $$k^j\partial_j~\equiv ~u\delta\partial_u+v_\ot\partial_{v_\ot}+v_\phi\partial_{v_\phi}
 +v_z\partial_{v_z}~\equiv~ \partial_R~,\eqno(6.11)$$

where $R=R(u)$ and so as usual 
$$\eqalignno{u&={e^{R\delta}\over\delta}sgn(\delta),&(6.12)\cr
{dR\over du}&=e^{-R\delta}sgn(\delta).&(6.13)\cr}$$

Proceeding exactly as before to solve for the invariants $X^{(i)}$ from 
$k^j\partial_jX^{(i)}=0$ gives the convenient choices 
$$\eqalignno{X^{(1)}&\equiv e^{-R}\,v_\ot,\cr
X^{(2)}&\equiv e^{-R}\, v_\phi,&(6.14)\cr 
X^{(3)}&\equiv e^{-R}\, v_z.\cr}$$
Moreover starting once again with the dimension space covector ${\bf a} 
\equiv (\delta,\nu,\mu)$, preserving $G$ (but not any characteristic length)
 and finding the dimension vectors of $f_1$, $F_2$ and $\Phi$ in the reduced 
 dimension space gives 
 $$\eqalignno{f_1~=~&\overline f_1(X^{(i)})\,e^{-(2\delta+\nu)R},\cr
 F_2&~=~\overline F_2(X^{(i)})\, e^{-(4\delta+3\nu)R},&(6.15)\cr
 \Phi&~=~\overline\Phi(X^{(i)})\, e^{2\nu R}.\cr}$$
 
We find subsequently that $\nu=1$ in accordance with the form assumed above 
for ${\bf k}$.
To proceed we first observe that by the scalings (6.14) and (6.15)  

 $$E~=~\left({X^{(j)}X_{(j)}\over 2}+\overline\Phi\right)e^{2R}\equiv 
 \overline E(X^{(j)})e^{2R}.\eqno(6.16)$$

 Then the compatibility of equations (6.8) and (6.15) (together with the isotropic 
 character of $f_1$ in velocity space) requires 

 $$f_1(E)=\overline f_1(E e^{-2R}) e^{-(2\delta +1)R}~.\eqno(6.17)$$

  It is clear therefore that $\overline f_1(\overline E)$ should be an 
  homogeneous function of order $\alpha$ (i.e. 
  $\overline f_1(kx)=k^\alpha \overline f_1(x)$) where 

  $$\alpha  =-(\delta +1/2).\eqno(6.18)$$

  In one dimension such a homogeneous function is a power law of power $\alpha$
   so that $\overline f_1(x)=Ax^\alpha$ and consequently 

$$f_1(E)=A\, E^\alpha.\eqno(6.19)$$

An exactly similar argument based on the compatibility of equations (6.9) 
   and (6.15) shows that 
   $$F_2(E)=B\, E^\beta,\eqno(6.20)$$
   where 
   $$\beta =-(2\delta+3/2).\eqno(6.21)$$
   In the preceeding formulae, $A$ and $B$ are arbitrary constants and $E$ should 
   be interpreted as the modulus of $E$ when $E<0$.
   
   Thus imposing a strict scaling symmetry (which will of course be communicated 
   to the particle orbits through the Hamiltonian by virtue of the scaling in 
   equation (6.16)) leads in a self-contained fashion to the distribution 
   function 
   $$\eqalignno{f&~=~\overline f_1(X^{(j)})\,e^{-(2\delta+1)R}+ j_z^2\, 
   \overline F_2(X^{(j)})\,e^{-(4\delta+3)R},&(6.22)\cr
    &~=~ A\, E^\alpha + B\,j_z^2\, E^\beta~ .&\cr}$$
   When proper account is taken of the scaling law of specific angular momentum ($
   \propto e^{-(\delta+1)R}$ by the usual dimension space arguments) equation 
   (6.21) shows explicitly that both terms in equation (6.22) scale in the 
   same way, as they should in order that $f$ have this same strict symmetry.
    
   We note that the potential varies as 
   $\Phi\propto u^{2/\delta}$, which on comparing with equation (2.1) of 
   E94 shows that his parameter $\beta_E$ is related to our 
   scaling parameter $\delta$ by $\beta_E=-(2/\delta)$. Our result (6.22) is then 
   equivalent to the form (2.6) of EdZ when the core 
   radius is zero, as it should strictly be.
   
   All of the discussion of this case is now reduced to the quadratures in 
   equations (6.6) which, together with (6.22) and the 
   scaling symmetry of equations (6.15), become 
   $$\eqalignno{4\pi\,G\,A \int\,\overline E^\alpha\,d\,X^{(1)}\,d\,X^{(2)}\,
   d\,X^{(3)}\,&~=~\overline\Phi\left(4\delta+{2(2-\delta)\over q^2}\right)~,&(6.23)\cr 
   4\pi\,G\,B\int\,\overline E^\beta\,(X^{(2)})^2\,d\,X^{(1)}\,d\,X^{(2)}\,
   d\,X^{(3)}\,&~=~4\overline\Phi\,\left({1\over q^2}-1\right)\,\delta^2(\delta-1)
   .&(6.24)\cr}$$   
      
From equation (6.23) one observes that for bound solutions $\overline\Phi<0$, 
we require 
$$\delta<sgn(A){1\over 1/2-q^2}~,\eqno(6.25)$$
and by equation (6.24) 
$$\left({1\over q^2}-1\right)(\delta-1)sgn(B)~<~0~.\eqno(6.26)$$
If both $A$ and $B$ are presumed positive then there is no cutoff scale except 
the zero energy surface. However there seems to be no need for this to be true,
and in general one can have an energy cutoff for any $j_z$ and vice versa.
 For unbound solutions with $E$ and $\overline\Phi$ both positive, we note that 
 $\delta$ should be positive if the gravitational acceleration is to be directed 
 inwards.        
 
 The evaluation of the integrals (6.24) and (6.25) has been extensively discussed 
 by E94 and EdZ and are easily done by changing variables to $\overline E$,
$\theta=\tan^{-1}\left (\left (v_\ot^2+v_\phi^2\right )^{1/2}/v_z\right )$,
and $\phi=\tan^{-1}\left (v_\phi/v_\ot\right )$.  In addition, one can calculate
various moments of the distribution function and for example, derive 
theoretical line profiles for stellar absorption line spectra.
 
\bigskip\bigskip

\line{\bf 7~~Conclusions\hfill}

\bigskip

In this article we have introduced a systematic method of imposing scaling 
symmetries or self-similarity in phase space.  These symmetries, along with various geometric
symmetries, allow one to find a wide range of analytic solutions to the
coupled Vlasov and Poisson equations.  This has led us 
to recover in a straightforward way the known `power law' or scaling solutions,
some of which are just beginning to be recognized for their practical 
implications.  In addition, we have found a new and very general class of 
spherical systems with velocity-space anisotroy.
The method is sufficiently powerful that we expect it to 
be useful in time dependent cases as well.  At the very least we will be able 
to reduce these problems by one variable as was achieved here for example in 
section 5.  
 
The self-similar solutions discussed in this paper
have applications in a wide variety of astrophysical problems.
For example, the solutions found in sections 3 and 4
provide examples of distribution functions where the
velocity dispersion increases without limit as $r\rightarrow 0$ (as per equation 2.23).   
These models might be relevant to observations of galactic nuclei where the densities are 
apparently correspondingly `cusped' (Tremaine et al. 1994). Although this might 
be thought to cast doubt on various black hole `detections', it must be remembered 
that the collisionless constraint has less and less relevance on galactic 
time scales in these dense regions. Thus the system will certainly evolve 
and be non-linearly unstable under the influence of these collisions. Nevertheless 
it is conceivable that recently formed structures might show this behaviour.
One application where the system is almost certainly collisionless,
even over a Hubble time, is the collapse of the dark matter.
Here, collapse can occur along one, two, or three axis and so,
following Fillmore and Goldreich (1984) we have studied self-similar
collapse in planar and cylindrical geometries. 

In section 5 we have found the power-law solutions that are
spherically symmetric though anisotropic in velocity space.
A subset of these models are the `power-law galaxies' of E94.  However,
our models are more general as illustrated in Figures 2(a)-2(d).  The
models therefore allow for more freedom in modelling galaxies and
fitting observables such as absorption-line spectra to theoretical
predictions.
    
Finally section 6 has enabled us to demonstrate the ease with which the 
spheroidal solutions may be found when the spatial symmetry and the scaling 
symmetry are imposed separately. Our conclusions are the same as those of 
E94 and of EdZ94, although we do point out that the core radius is not actually 
playing a role in these solutions and it is apparent that these solutions 
form part of the wider family of solutions discussed in this paper.
We also remark that hyperboloidal solutions of this type obviously also exist.

Future work will focus on the time dependent solutions with and without an 
expanding background partly in hopes of obtaining the results of Fillmore and Goldreich 
(ibid) without the singularities in the density, but also in order to make a survey 
similar to the present survey of the stationary solutions. More detailed investigation 
of the stability and other properties of the solution in section 5 is also 
required.

\bigskip
\bigskip

\line{\bf{Acknowledgements}\hfill}

\bigskip

It is a pleasure to acknowledge helpful conversations with Drs W-Y Chau, Martin 
Duncan, and Scott Tremaine.   
 
\bigskip

\vfill

\eject

\bigskip

\line{\bf References\hfill}

\bigskip

\noindent Aly J.-J., 1989, MNRAS {\bf 241}, 15

\bigskip

\noindent Aly J.-J., 1994, "Ergodic Concepts in Stellar Dynamics", 
V.G. Gurzadyan and D Pfenniger (eds), Springer-Verlag, p226

\bigskip

\noindent Barenblatt G.E. and Ya. B.  Zel'dovich 1972, Ann.Rev.Fluid Mech.
{\bf 4}, 285 (BZ)

\bigskip

\noindent Bertschinger E., 1985, ApJS, {\bf 58}, 39

\bigskip

\noindent Binney J., Tremaine S., 1987, Galactic Dynamics. Princeton Univ. Press,
Princeton, NJ, Ch. 4 (BT)
             
\bigskip

\noindent Carter B., Henriksen R. N., 1991, J. Math. Phys. , 
{\bf 32}, 2580 (CH)

\bigskip

\noindent Evans N. W., 1994, MNRAS, {\bf 267}, 333 (E94)

\bigskip

\noindent Evans N. W., de Zeeuw P. T., 1994, MNRAS, {\bf 271}, 202 (EdZ)

\bigskip

\noindent Fillmore J. A., Goldreich P., 1984, ApJ, {\bf 281}, 1

\bigskip

\noindent Fridman A. M., Polyachenko V. L., 1984, Physics of Gravitating
Systems I. Springer-Verlag, NY, Ch. 3
             
\bigskip

\noindent Fujiwara T., 1983, PASJ, {\bf 35}, 547

\bigskip

\noindent Gurevich A. V., Zybin K. P., 1988, Zh. Eksp. Teor. Fiz. , {\bf 94}, 3

\bigskip

\noindent Gurevich A. V., Zybin K. P., 1990, Sov. Phys. JETP, {\bf 70}, 10

\bigskip

\noindent Lynden-Bell, 1967, MNRAS, {\bf 136}, 101

\bigskip 

\noindent Lynden-Bell D., Eggleton P. P., 1980, MNRAS, {\bf 191}, 483

\bigskip
             
\noindent Inagaki S., Lynden-Bell D., 1983, MNRAS, {\bf 205}, 913

\bigskip

\noindent Inagaki S., Lynden-Bell D., 1990, MNRAS, {\bf 244}, 254

\bigskip

\noindent Ryden B. 1993, ApJ {\bf 418}, 4

\bigskip

\noindent Tremaine S., H\'enon M. and Lynden-Bell D., 1986, MNRAS, {\bf 219}, 285
\bigskip

\noindent Tremaine S., Richstone D. O., Byun Yong-Ik, Dressler A., Faber S. M.,
Grillmair C., Kormendy J., and Lauer T. R., 1994, AJ {\bf 107}, 634

\bigskip

\noindent van der Marel R. P., Franx M., 1993, ApJ, {\bf 407}, 525

\bigskip

\noindent Wiechen,H., Ziegler, H.J. and Schindler,K., 1988, MNRAS
 {\bf 232}, 623

\vfill
\eject

\line{\bf Figure Captions\hfill}

\bigskip

{\bf Figure 1.}~~Contour plot of the distribution function for a spherically
symmetric system with radial orbits.  The distribution function is given by
equations (2.19) and (2.20) with $\delta=-17$ (chosen so that this figure corresponds 
to Fig. 10 of Fillmore and Goldreich (1984)).  Horizontal and vertical axis are
$r$ (measured in units of some fiducial length `$a$'
and $v_r$ (measured in units of $\Phi_a^{1/2}$).
Contours are linearly spaced and range from $0.04$ to $1.0$.

\bigskip

{\bf Figure 2(a).}~~Contour plot of the distribution function for a system with that is
isotropic in velocity space ($\beta=0$).  
The distribution function is given by equation (5.21)
or equivalently (5.20) with $F(\xi)=\xi^\alpha$.  For definiteness,
we take $\delta=-8$ and $\alpha=-1.07$.
$r$ is fixed and the horizontal and vertical
axis are $\left ({v_\theta^2}+{v_\phi^2}
\right )^{1/2}= j/r$ and $v_r$ measured in units of $\Phi(r)^{1/2}$.
The contours are logarithmic ranging from $1$ to $10^5$ with
the distribution function increasing towards the lower left of the
plot.  (We have not worried about 
normalization of the distribution function.)

\bigskip

{\bf Figure 2(b).}~~Same as Fig. 2(a) but for a distribution function 
constructed from nearly radial orbits ($\beta=0.9,~\delta=-8.,~\alpha=
-0.17$).  

\bigskip

{\bf Figure 2(c).}~~Same as Fig. 2(a) but for a distribution function 
constructed from nearly circular orbits ($\beta=-9.0,~\delta=-8.,~\alpha=
-10.07$).   

\bigskip

{\bf Figure 2(d).}~~Same as Fig. 2(d) but here the distribution function
is given by the sum of the two power-law distributions used in Figs
2(b) and 2(c).  By construction, this distribution has $\beta=0$.

\bigskip
 
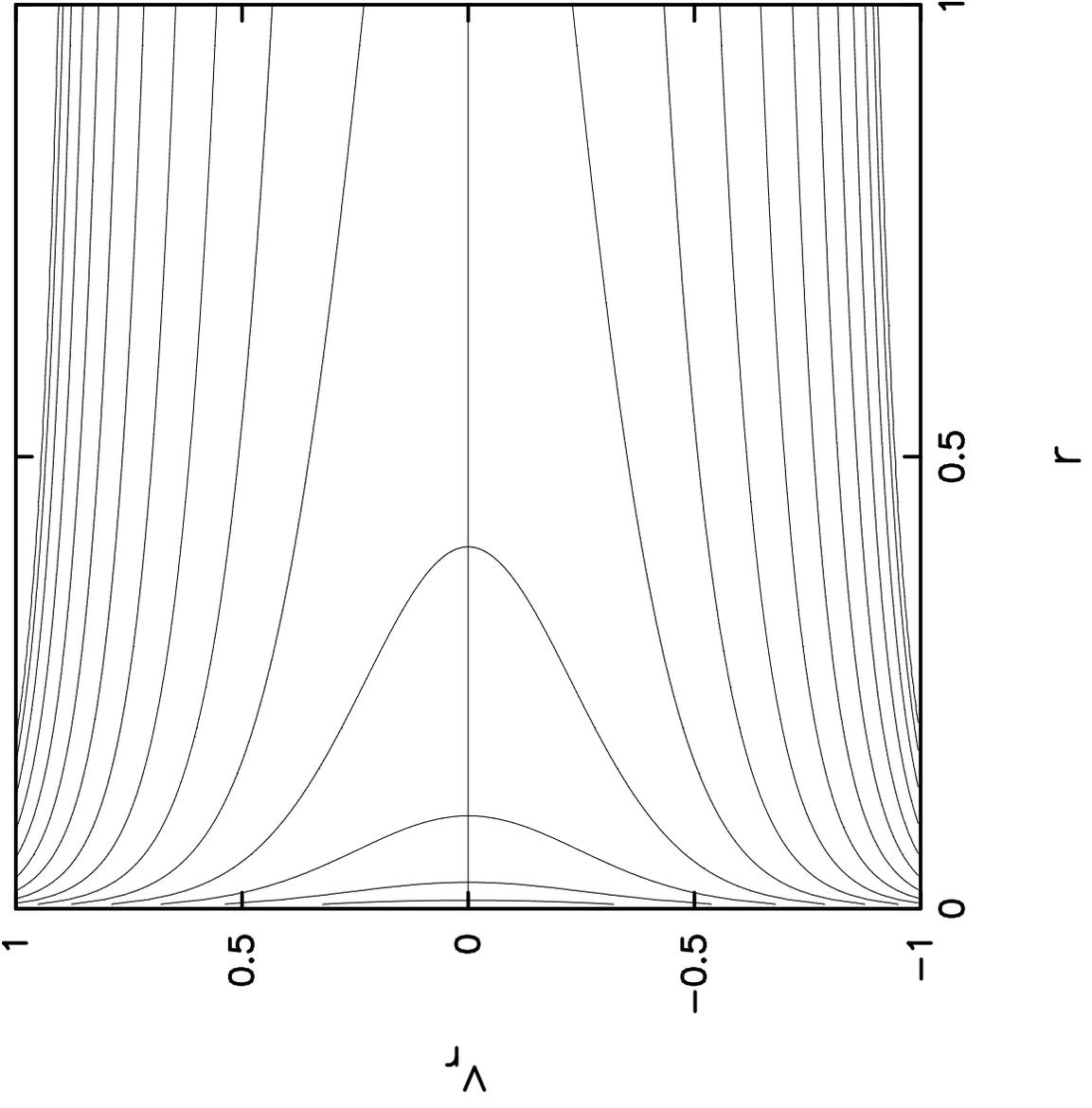
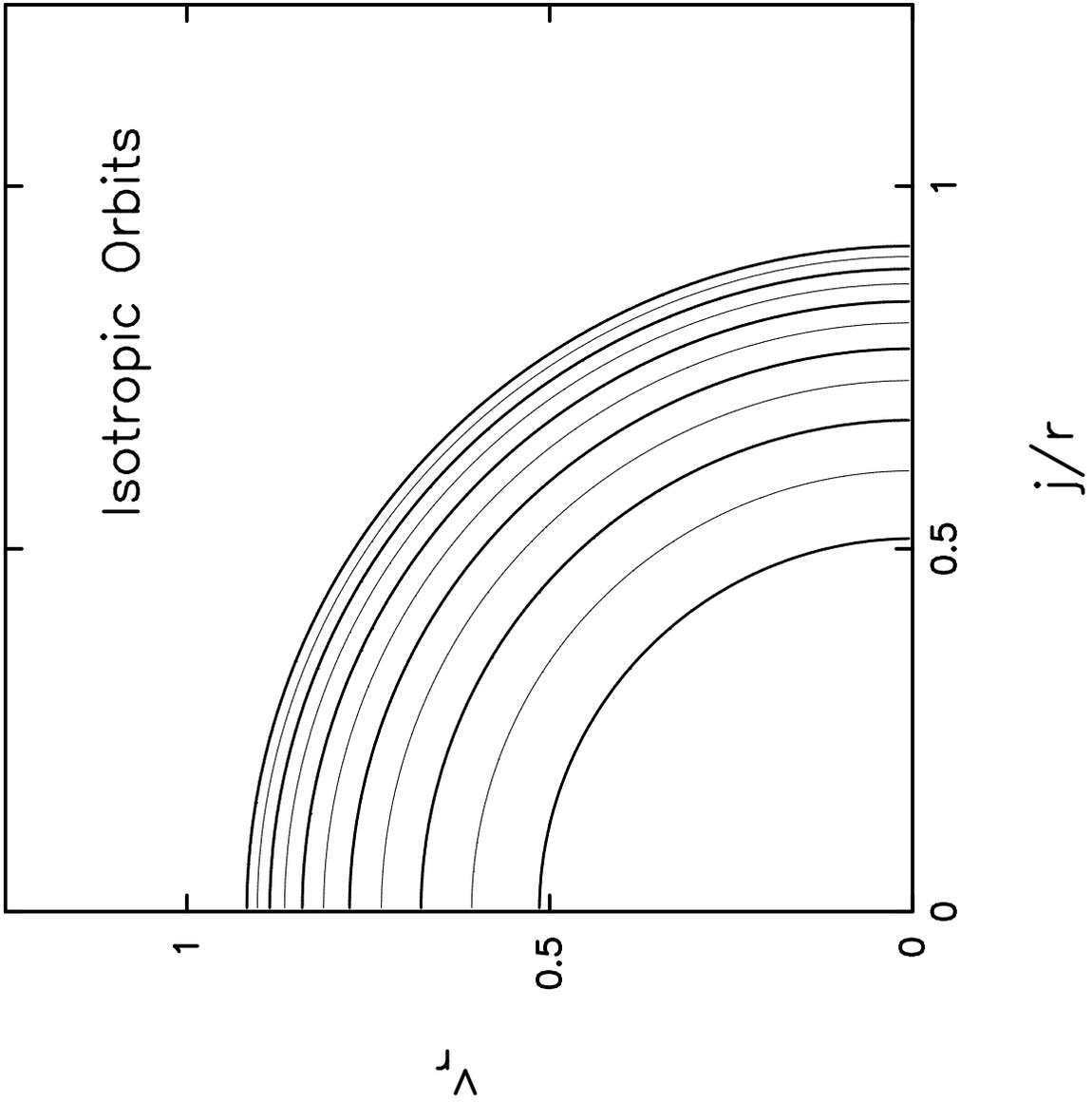
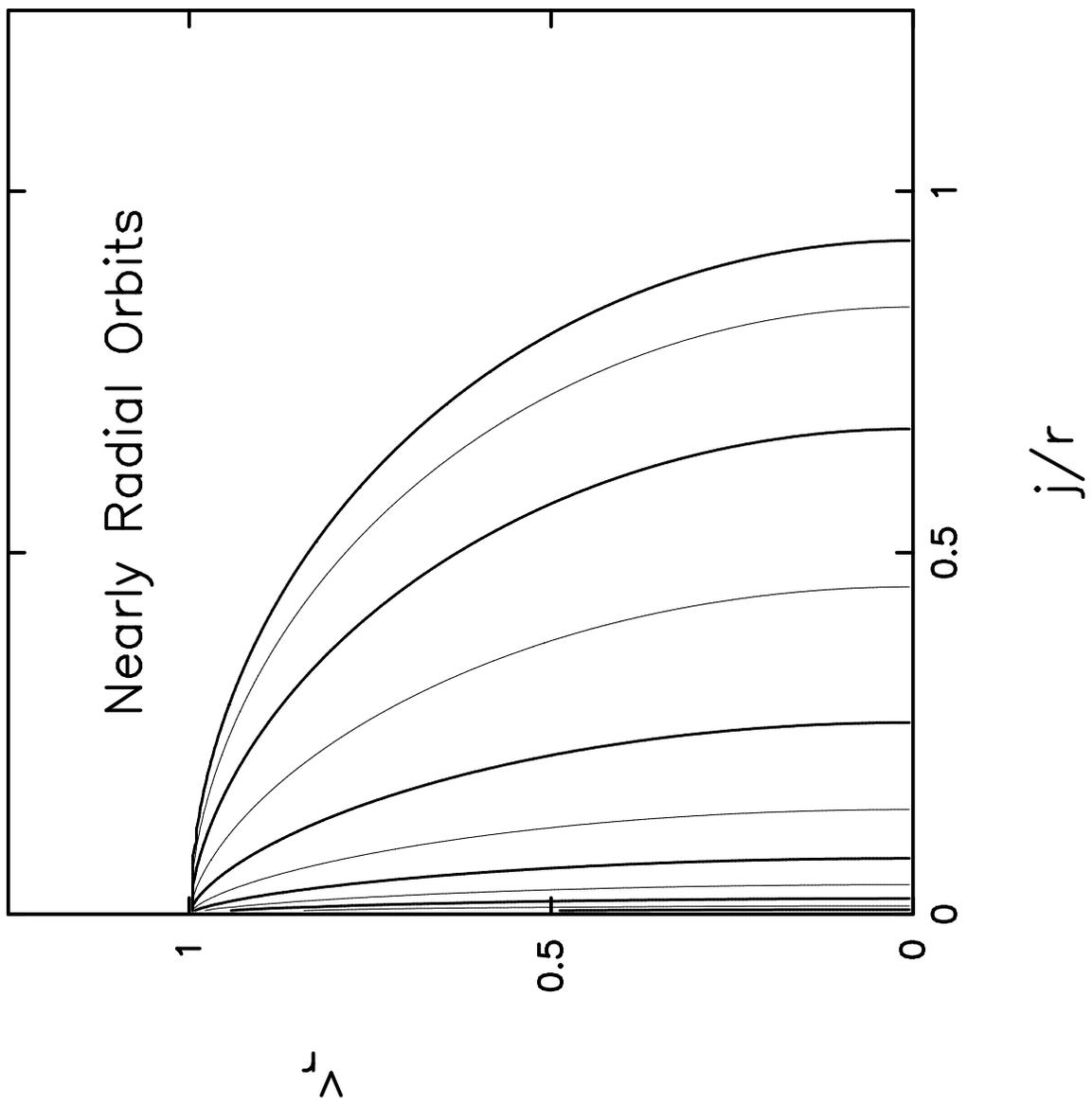
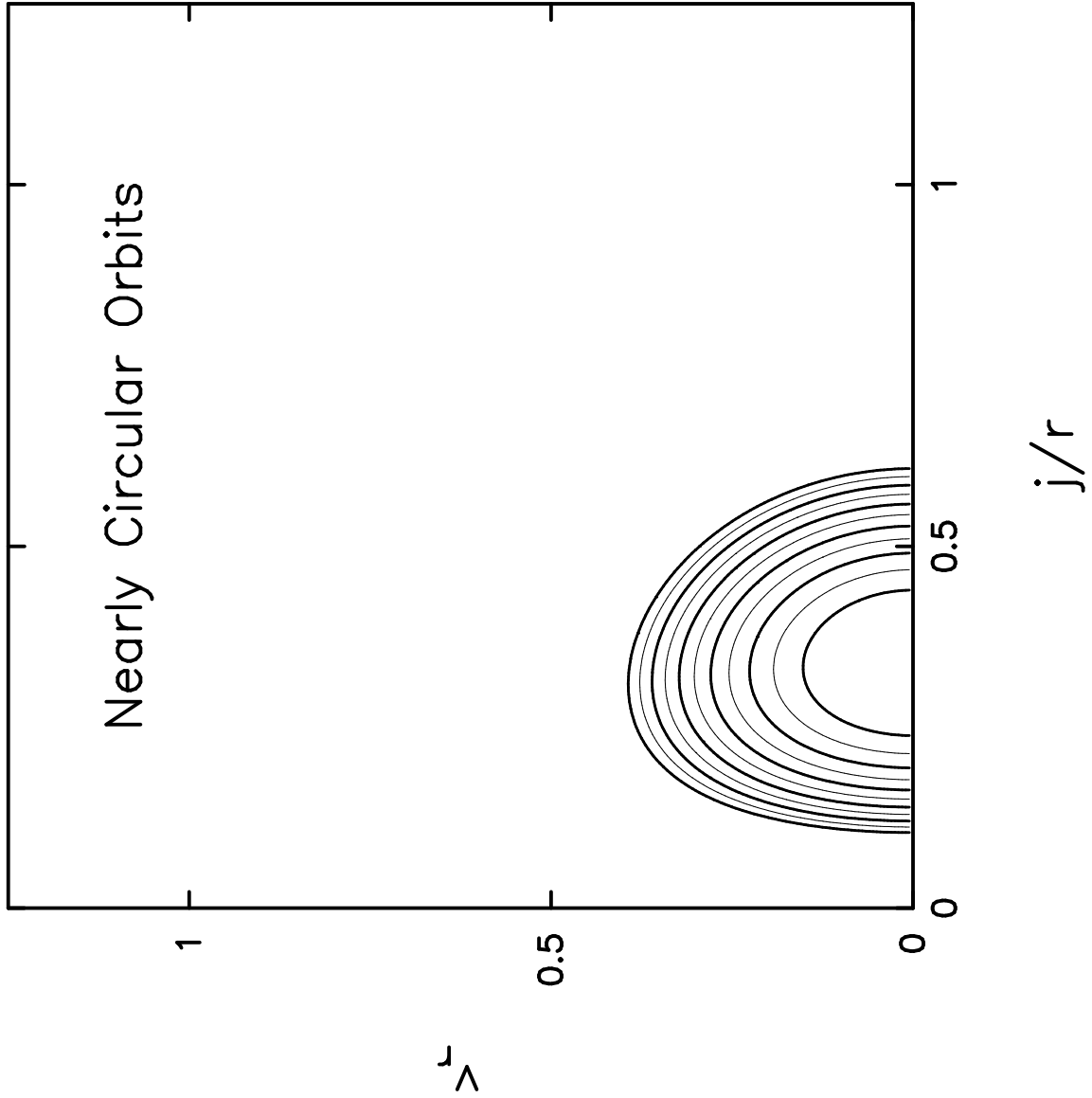
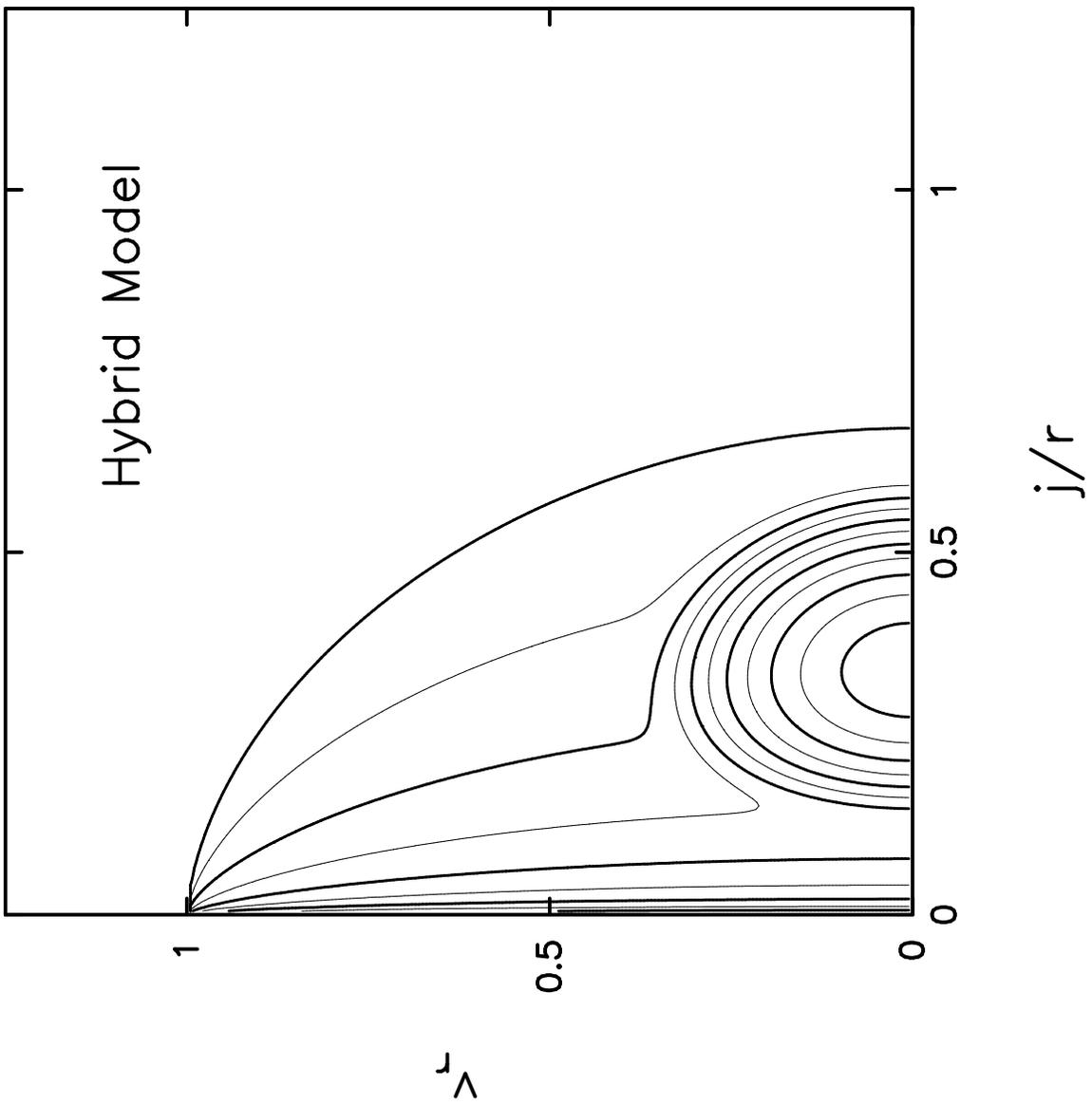
\bye